\documentclass[preprint,showpacs,preprintnumbers,amsmath,amssymb]{revtex4}
\usepackage{epsfig}

\begin{document}

\title{Twisted phonons in Bose Einstein condensates}

\author{J.T. Mendon\c{c}a}

\email{titomend@ist.utl.pt}

\affiliation{Instituto de F\'isica, Universidade de S\~ao Paulo, S\~ao Paulo SP, 05508-090 Brasil
\\ IPFN, Instituto Superior T\'ecnico, 1049-001 Lisboa, Portugal}

\author{A. Gammal}

\email{gammal@if.usp.br}

\affiliation{Instituto de F\'isica, Universidade de S\~ao Paulo, S\~ao Paulo SP, 05508-090 Brasil
\\ IPFN, Instituto Superior T\'ecnico, 1049-001 Lisboa, Portugal}

\begin{abstract}

We consider elementary excitations in a Bose Einstein condensate, carrying a finite amount of angular momentum. We show that these elementary excitations are modified Bogoliubov oscillations, or phonons with an helical wave structure. These twisted phonon modes can contribute to the total vorticity in a quantum fluid, thus complementing the contribution of the traditional quantum vortices. Linear and nonlinear versions of twisted phonon modes will be discussed. New envelope soliton solutions are shown to exist in a condensate. 

\end{abstract}

\maketitle

\section{Introduction}
Vorticity has been a central problem in Bose Einstein condensates and superfluids, as discussed in several review papers \cite{fetter,sonin} and books \cite{pitaevskii,leg,phetick,book2}. Rotating condensates with vortical structures have also been studied experimentally by many groups \cite{dalibard,ketterle,cornell}. Furthermore, vortex structures are considered a central piece of the turbulence state in quantum fluids.
According to the traditional view, turbulence in a condensate is made of a superposition of vortex structures and phonon modes. Vortices are local structures of quantum origin, which can occur around regions of phase singularity. Phonons are elementary excitations of the condensed fluid which, in principle, carry no finite vorticity. However, as shown here, these elementary excitations can also carry a finite amount of vorticity, and can be seen as a new kind of propagating vortical structures. 

Twisted phonons, or acoustic waves with a finite amount of angular momentum, have already been considered in classical fluids \cite{zhang,thomas}. It is the purpose of the present work to show that similar modes can also be considered in a quantum fluid. We will consider both linear and nonlinear phonon mode solutions in a condensate, carrying a finite angular momentum. In particular, we will show that a new kind of soliton solutions exist in a condensate, which are distinct from the usual soliton solutions of the Gross-Pitaevskii (GP) equation. In contrast with the usual view of solitons in a condensate, they are envelope solitons associated with elementary oscillations. In that sense, they can be seen as the nonlinear version of the twisted phonons, and can also carry a finite amount of angular momentum.

The structure of the paper is the following. In Section II, we state our basic equations and derive a nonlinear wave equation for phonons in a condensate, which will be used as the starting point of our calculations. In section III, we consider a linear mode analysis which will allow us to establish the space and time structure of twisted Bogoliubov oscillations. We will show that the linear wave solutions can be described as a superposition of Laguerre-Gauss modes, similar to those already considered in classical fluids. Each of these modes carry a finite amount of angular momentum, as shown in Section IV. These results can be extended to the nonlinear regime, as shown in Section V. Twisted envelope solitons, which are different from the traditional soliton solutions considered in condensates \cite{phetick,book2} are shown to exist. They are described by a defocusing nonlinear Shr\"odinger (NLS) equation for the phonon mode amplitude. Finally, in Section VI, we state some conclusions.

\section{Basic equations}

We start with the GP equation, which is formally identical to a NLS equation, and adequately describes the condensate in the mean field approximation. It can be written as
\begin{equation}
i \hbar \frac{\partial}{\partial t} \psi = \left[ - \frac{\hbar^2 \nabla^2}{2 M} + V_{ext} + g \left| \psi \right|^2 \right] \psi \, ,
\label{2.1} \end{equation}
where $V_{ext}$ is an external confining potential, $M$ is the mass of the condensed atoms, $g$ is the coupling parameter resulting from binary atomic collisions in the zero energy limit, and $\psi \equiv \psi ({\bf r}, t)$ is the wave function of the condensed matter. Using the Madelung transformation 
\begin{equation}
\psi ({\bf r}, t) = \sqrt{n} \; \exp{( i \phi )} \, ,
\label{2.2} \end{equation}
where $n$ is the density and $\phi$ is the phase function, and defining the fluid velocity as ${\bf v} = \hbar \nabla \phi / M$, we can derive from (\ref{2.1}) the quantum fluid equations, as
\begin{equation}
\frac{\partial n}{\partial t} + \nabla (n {\bf v} ) = 0 \, ,
\nonumber \end{equation}
\begin{equation}
\frac{\partial {\bf v}}{\partial t} + {\bf v} \cdot \nabla {\bf v} = -\frac{\nabla P}{M n} - \frac{1}{M} \nabla (V_{ext} +V_B ) \, ,
\label{2.3} \end{equation}
where the pressure $P$ results from the atomic collisions, and the Bohm potential $V_B$ describes quantum non-locality, are defined by
\begin{equation}
P = \frac{1}{2} g n^2 \; , \quad V_B = - \frac{\hbar^2}{2 M} \frac{\nabla^2 \sqrt{n}}{\sqrt{n}} \, .
\label{2.4} \end{equation}
The quantity $- \nabla V_B$ is also usually called the quantum pressure. Taking the time derivative of the first of eqs. (\ref{2.3}), and using vectorial identities, we can derive an evolution equation of the form
\begin{equation}
\frac{\partial^2 n}{\partial t^2} = \nabla^2 \left( n v^2 + \frac{1}{2} g n^2 \right) + \frac{1}{M} \nabla \cdot \left[ n \nabla (V_{ext} +V_B ) \right] \, .
\label{2.5} \end{equation}
This will be the basic equation of our present model. Up to this point, no approximations have been made. The linear and nonlinear vortex solutions of this equation will be considered successively. Our analysis will be able to show that phonon vortices can be excited in a condensate. They are intrinsically different from the well known quantum vortex solutions, in the sense that they are space-time structures propagating in the bulk of the condensed gas, and can be described by a linear model, as shown next. 

\section{Twisted phonon modes}

Let us first consider a small density perturbation $\tilde n$, such that $| \tilde n | \ll n_0$, where $n_0$ is the equilibrium density. Retaining only the linear terms in (\ref{2.5}), we obtain
\begin{equation}
\frac{\partial^2 \tilde n}{\partial t^2} - \frac{n_0}{M} \nabla^2 ( g \tilde n + \tilde V_B ) = 0 \, ,
\label{3.1} \end{equation}
with the perturbed Bohm potential
\begin{equation}
\tilde V_B = - \frac{\hbar^2}{4 M} \frac{\nabla^2 \tilde n}{n_0} \, .
\label{3.1b} \end{equation}
The perturbations associated with the external potential $V_{ext}$ are ignored here. This is justified for small scale elementary excitations in the condensate, which are assumed much smaller than the size of the condensed cloud. At this scale, the equilibrium density can be assume flat, in contrast with the analysis of global modes, where density profiles and boundary conditions become an essential ingredient \cite{stringari,hugo}.

Let us then assume propagation of an acoustic mode along some arbitrary direction $Oz$, as described by a solution of the form
\begin{equation}
\tilde n ({\bf r}, t) = a ({\bf r}) \exp \left( i k z - i \omega t \right) \, ,
\label{3.2} \end{equation}
where $\omega$ is the mode frequency, and $a ({\bf r})$ is a slowly varying amplitude satisfying the paraxial approximation
\begin{equation}
\left| \frac{\partial^2 a}{\partial z^2} \right| \ll \left| k \frac{\partial a}{\partial z} \right| \, .
\label{3.2b} \end{equation}
Replacing this in the linearized equation (\ref{3.1}), we get
\begin{equation}
\left( c_s^2 \nabla_p^2 + \frac{\hbar^2}{4 M} \nabla_p^4 \right) a = 0 \, ,
\label{3.3} \end{equation}
where
\begin{equation}
\nabla_p^2 = \nabla_\perp^2 + 2 i k \frac{\partial}{\partial z} \, ,
\label{3.3b} \end{equation}
is the paraxial operator, and $c_s = \sqrt{g n_0 / M}$ is the Bogoliubov sound speed. In the derivation of eq. (\ref{3.3}) we have assumed that the frequency $\omega$ satisfies the quantum phonon dispersion 
\begin{equation}
\omega^2 = c_s k^2 + \frac{\hbar^2 k^4}{4 M^2} \, .
\label{3.4} \end{equation}
In order to solve the evolution of the mode amplitude $a ({\bf r})$, we notice that eq. (\ref{3.3}) is satisfied if the following reduced equation is also satisfied
\begin{equation}
\nabla_p^2 \; a \equiv \left(\nabla_\perp^2 + 2 i \frac{\partial}{\partial z} \right) a = 0
\label{3.5} \end{equation}
This is the well known paraxial equation describing the evolution of a wave beam near the focal region. Although the solutions of eq. (\ref{3.5}) also satisfy the generalized paraxial equation (\ref{3.3}), the reverse is not necessarily true. In the following, we will focus our analysis on the solutions of this simple paraxial equation. It is well known that any solution of eq. (\ref{3.5}) can be represented as a superposition of Laguerre-Gauss (LG) modes, of the form
\begin{equation}
a ({\bf r}) = \sum_{pl} a_{pl} F_{pl} ({\bf r}) \, ,
\label{3.6} \end{equation}
where $a_{pl}$ are the amplitudes, the integers $p$ and $l$ are the radial and azimuthal quantum numbers, and the LG basic functions are defined by
\begin{equation}
F_{pl} (r, \theta, z) = C_{pl} L_p^{|l|} (X) X^{|l|} e^{-X/2} e^{i l \theta} \, ,
\label{3.7} \end{equation}
where cylindrical coordinates were used, ${\bf r}_\perp \equiv (r, \theta)$, $L_p^l$ are the associated Laguerre polynomials and $C_{pl}$ are normalization constants, and defined by
\begin{equation}
C_{pl} = \frac{1}{2 \sqrt{\pi}} \left[ \frac{(l + p)!}{p!} \right]^{1/2} \, , \quad
L_p^l (X) = \frac{e^X X^{- l}}{l!} \frac{d^p}{d X^p} \left( e^{- X} X^{l + p} \right) \, .
\label{3.8} \end{equation}
It is important to notice that the LG modes (\ref{3.7}) satisfy the orthogonality relations
\begin{equation}
\int_0^\infty r dr \int_0^{2 \pi} d \theta \; F_{pl}^* F_{p'l'} = \delta_{p p'} \delta_{l l'} \, .
\label{3.9} \end{equation}
We have also used an auxiliary variable $X = r^2 / w^2$, where the beam waist $w \equiv w (z)$ is a slowly varying function of the axial variable $z$.  Each of these $(p, l)$ modes represents a phonon vortex structure, which can be called a twisted phonon mode. These modes carry a finite amount of angular momentum, as explicitly considered below. It is important to notice that the surfaces of constant phase of these twisted modes are not planes as in the usual phonon mode structure, but are helical surfaces instead, as determined by the condition $\varphi \equiv k z + l \theta = const.$

\section{Phonon vorticity}

We now consider the energy, momentum and angular momentum carried by each LG mode. We start by considering the linear momentum density, as defined by ${\bf G} = M n {\bf v}$. The average momentum associated with the LG mode spectrum inside the condensate can then be determined by
\begin{equation}
{\bf G} = \frac{1}{2} M \sum_{pl} \left( \tilde n_{pl} {\bf v}_{pl}^* +  \tilde n_{pl}^* {\bf v}_{pl} \right) \, .
\label{4.1} \end{equation}
Using the continuity equation, we can determine the velocity perturbations associated to the LG phonon modes, as given by $\omega \tilde n_{pl} = n_0 (i \nabla_\perp + k {\bf e}_z ) \cdot {\bf v}_{pl}$, enabling us to write
\begin{equation}
{\bf v}_{pl} = \frac{\omega}{q^2} \frac{\tilde n_{pl}}{n_0} {\bf q} \, ,
\label{4.2} \end{equation}
where the wavevector ${\bf q}$ is perpendicular to the helical surfaces of constant phase $\varphi \equiv k z + l \theta = const.$ We therefore get ${\bf q} = \nabla \varphi$, or 
\begin{equation}
{\bf q} = k \, {\bf e}_z + \frac{l}{r} \, {\bf e}_\theta \, .
\label{4.3} \end{equation}
A small radial component could also be added, but is ignored here for simplicity. Replacing this in eq. (\ref{4.1}), and noting that in the paraxial approximation we have $| {\bf q} | \simeq k$, we can write for the modes with frequency $\omega$, the expression
\begin{equation}
{\bf G}  = \frac{M}{n_0} \left( \frac{\omega}{k^2} \right) \sum_{pl} \left| \tilde n_{pl} \right|^2 {\bf q} \, .
\label{4.4} \end{equation}
We can now define the angular momentum density of the phonon spectrum as
${\bf M} = \left( {\bf r} \times {\bf G} \right)$. As a result, the average angular momentum associated with the phonon modes can be determined by
\begin{equation}
{\bf M} = \frac{1}{2} M \sum_{pl} \left[ \tilde n_{pl} ( {\bf r} \times {\bf v}_{pl}^*) +  \tilde n_{pl}^* ({\bf r} \times {\bf v}_{pl} )\right] \, .
\label{4.5} \end{equation}
Using the above results, we can then establish the axial component of the angular momentum $M_z$ associated with phonon modes at the frequency $\omega$, as
\begin{equation}
M_z  = \frac{M}{n_0} \left( \frac{\omega}{k^2} \right) \sum_{pl} l \left| \tilde n_{pl} \right|^2 \, .
\label{4.6} \end{equation}
This shows that each LG mode carries an axial angular momentum given by
\begin{equation}
M_z (p, l) = l f (p, l) \; , \quad f (p, l) =  \frac{M}{n_0} \left( \frac{\omega}{k^2} \right) \left| \tilde l n_{pl} \right|^2
\label{4.7} \end{equation}
Finally, we can define the energy density, following \cite{zhang}, as
\begin{equation}
\mathcal{E} = \frac{1}{2} M n_0 v^2 + \frac{1}{2} \frac{P^2}{M n_0 c_s^2}
\label{4.8} \end{equation}
Using the velocity perturbations (\ref{4.2}), and considering average pressure perturbations of the form
\begin{equation}
\tilde P = \frac{1}{2} g n_0 \tilde n = \frac{1}{2} M c_s^2 \tilde n
\label{4.9} \end{equation}
we establish the average value of the energy density as
\begin{equation}
\mathcal{E} = \frac{M}{n_0} \left[ \left( \frac{\omega}{k} \right)^2 + \frac{1}{4} c_s^2 \right] \sum_{pl} l \left| \tilde n_{pl} \right|^2 \, .
\label{4.10} \end{equation}
Noting that $c_s^2 \simeq (\omega / k)^2$, we can simplify this expression to
\begin{equation}
\mathcal{E} \simeq \frac{M}{n_0} \left( \frac{\omega}{k} \right)^2  \sum_{pl} l \left| \tilde n_{pl} \right|^2 \, .
\label{4.10b} \end{equation}
It then becomes obvious that the energy density for a given LG mode can be determined by
\begin{equation}
\mathcal{E} (p, l) = \omega f(p, l) \, .
\label{4.11} \end{equation}
From here we obtain a simple relation between the energy and the axial angular momentum of the twisted phonon mode, as
\begin{equation}
M_z (p, l) = \frac{l}{\omega} \mathcal{E} (p, l) \, .
\label{4.11b} \end{equation}
A similar relation exists between the energy and the axial component of the linear momentum ${\bf G}$. Using (\ref{4.4}), we can write ${\bf G} = \sum_{pl} {\bf G} (p, l)$, with 
\begin{equation}
G_z (p, l) = k f (p, l) = \frac{k}{\omega} \mathcal{E} (p, l) \, .
\label{4.12} \end{equation}
Similar results have already been obtained for sound waves in classical fluids \cite{zhang,thomas}. These relations can easily be transposed to the quantum operator language. If we describe the spectrum of phonon perturbations as a boson field, with creation and destruction operators per mode $\hat b_{pl}^\dag$ and $\hat b_{pl}$, we can define the energy operator per mode as
\begin{equation}
\hat{\mathcal{E}} (p, l) = \hbar \omega \left( \hat N_{pl} + \frac{1}{2} \right) \, ,
\label{4.13} \end{equation}
with the phonon mode number operator  $\hat N_{pl} = \hat b_{pl}^\dag \hat b_{pl}$. The axial linear and angular momentum operators become 
\begin{equation}
\hat G_z (p, l) = \hbar k \left( \hat N_{pl} + \frac{1}{2} \right)  \; , \quad 
\hat M_z (p, l) = \hbar l \left( \hat N_{pl} + \frac{1}{2} \right) \, .
\label{4.13} \end{equation}
This shows that the axial linear and angular momenta of an elementary quantum LG excitation are equal to $\hbar k$ and $\hbar l$, respectively. The vorticity content of an LG phonon is therefore of the same order of the vorticity associated with the usual quantum vortices. This completes our discussion of the linear properties of the twisted phonon modes in a condensate. Next, we discuss the possible existence of nonlinear twisted phonon oscillations, in the form of exact soliton solutions. 

\section{Twisted solitons}

Let us then consider nonlinear wave solutions of our basic equation (\ref{2.5}). For this purpose, we expand it up to the third order in the density perturbation $\tilde n$. We obtain
\begin{equation}
\frac{\partial^2 \tilde n}{\partial t^2} - \frac{n_0}{M} \nabla^2 ( g \tilde n + \tilde V_B ) = \mathcal{NL} \, ,
\label{5.1} \end{equation}
where $\tilde V_B$ is still given by (\ref{3.1b}), and the second and third order nonlinear terms are represented by $\mathcal{NL} = \mathcal{NL}_{2nd} + \mathcal{NL}_{3rd}$, with
\begin{equation}
\mathcal{NL}_{2nd} = \left( n_0 \nabla^2 v^2 + \frac{1}{2} g \nabla^2 n^2 \right) - \frac{\hbar^2}{4 M^2} \left[ \frac{\tilde n}{n_0} \nabla^2 \tilde n - \frac{(\nabla \tilde)^2}{2 n_0} \right] \, ,
\label{5.1b} \end{equation}
and
\begin{equation}
\mathcal{NL}_{3rd} =  \nabla^2 ( \tilde n v^2 )  - \frac{\hbar^2}{4 M^2} \left[ \frac{\tilde n^2}{n_0^2} \nabla^2 \tilde n + \frac{2 \tilde n}{n_0^2} (\nabla \tilde)^2 \right] \, .
\label{5.1c} \end{equation}
We now focus on twisted wave solutions of the form (\ref{3.2}), but consider a single LG mode, as described by
\begin{equation}
\tilde n ({\bf r}, t) = a (z, t) F_{pl} ({\bf r}) \exp \left( - i \omega t \right) \, ,
\label{5.2} \end{equation}
where we assume that $a (z, t)$ is a slowly varying function of time, but also includes a fast scale spatial variation approximately given by $\exp (i k z)$, as specified later. Noting that only the third order terms of eq. (\ref{5.1c}) contribute to this mode, and that the second order terms of eq. (\ref{5.1b}) are only relevant to nonlinear wave mixing and not to single mode analysis, we obtain after integration over the perpendicular space variable ${\bf r}_\perp$, the following evolution equation
\begin{equation}
\left[ \omega^2 + 2 i \omega \frac{\partial}{\partial t} - \left( c_s^2 - \frac{\hbar^2}{4 M^2} \frac{\partial^2}{\partial z^2} \right) \frac{\partial^2}{\partial z^2} \right] a = k^2 R_{pl} \left( | {\bf v} |^2 + \frac{3 \hbar^2 k^2}{4 M^2} \frac{| a |^2}{n_0^2} \right) a \, , 
\label{5.3} \end{equation}
where we have defined 
\begin{equation}
R_{pl} = \int_0^\infty \left| F_{pl}^2 ({\bf r}) \right|^2 r dr \, .
\label{5.3b} \end{equation}
It can easily be realized that, in this expression, the integrand is independent of the angular variable $\theta$. Using the linear wave solutions for the velocity perturbations, as given by eq. (\ref{4.2}), and noting that $q \simeq k$, we get
\begin{equation}
| {\bf v} |^2 = \frac{\omega^2}{k^2} \frac{ | a |^2}{n_0^2} \, .
\label{5.4} \end{equation}
This allows us to rewrite the nonlinear mode equation (\ref{5.3}) in the following compact form
\begin{equation}
\left( \omega^2 + 2 i \omega \frac{\partial}{\partial t} -  \frac{D^2}{D z^2} \right) a + W (\omega) | a |^2 a = 0 \, , 
\label{5.5} \end{equation}
where we have introduced the notation
\begin{equation}
\frac{D^2}{D z^2} \equiv \left( c_s^2 - \frac{\hbar^2}{4 M^2} \frac{\partial^2}{\partial z^2} \right) \frac{\partial^2}{\partial z^2}  \, ,\label{5.5b} \end{equation}
and defined the nonlinear coupling coefficient $W (\omega)$, such that 
\begin{equation}
W (\omega) = \frac{R_{pl}}{n_0^2} \left( \omega^2 + \frac{3 \hbar^2 k^4}{4 M^2} \right) \, .
\label{5.6} \end{equation}
This shows an explicit contribution of quantum dispersion to the nonlinear coupling. At this point, the formal similarity of eq. (\ref{5.5}) to a nonlinear Schr\"odinger equation should be noticed. Such a similarity will be explored further. It is now useful to introduce a variable transformation, from the old space and time variables $(z, t)$ to the new variables $(\xi, \tau)$, such that
\begin{equation}
\xi = z - \frac{\omega}{k} t \; , \quad \tau = W (\omega) t \, .
\label{5.7} \end{equation}
With such a transformation, we can replace the amplitude function $a (z, t)$ by a solution of the form
\begin{equation}
a (\xi, \tau ) = A (\xi, \tau) \exp (i k z) \equiv A (\xi, \tau) \exp \left( i k \xi - i \frac{\omega}{W (\omega)} \tau \right) \, .
\label{5.8} \end{equation}
Replacing this in (\ref{5.5}), retaining the dominant terms of the operator (\ref{5.5b}) and using the linear dispersion relation (\ref{3.4}), we finally obtain for the slowly varying amplitude $A (\xi, \tau)$ the simple evolution equation
\begin{equation}
i \frac{\partial A}{\partial \tau} + \frac{\omega}{2 k^2 W (\omega)} \frac{\partial^2 A}{\partial \xi^2} - \frac{| A |^2}{2 \omega} A = 0 \, .
\label{5.9} \end{equation}
This is a defocusing NLS equation, similar to the initial GP equation (\ref{2.1}) for repulsive interactions, $g > 0$. However it has a very different physical meaning, because it concerns the amplitude of twisted phonon modes, and not the condensate wavefunction. Furthermore, it is insensitive to the nature of the atomic collisions, because the new nonlinear terms are independent of the sign of the interaction parameter $g$. In order to discuss the solutions of this equation pertinent to our problem, it is useful to introduce new dimensionless variables $(\eta, s)$, and a new amplitude function $u (\eta, s)$, as defined by
\begin{equation}
s =  \frac{\omega}{W (\omega)} \tau = \omega t \; , \quad \eta = k \xi = k z - \omega t \; , \quad 
A (\eta, s) = \omega \sqrt{\frac{2}{W (\omega)}} u (\eta, s) \, .
\label{5.10} \end{equation}
We can then rewrite eq. (\ref{5.9}) as
\begin{equation}
i \frac{\partial u}{\partial s} + \frac{1}{2} \frac{\partial^2 u}{\partial \eta^2} - | u |^2 u = 0 \, .
\label{5.11} \end{equation}
This is the defocusing NLS equation in its standard form. It is well known that such an equation satisfies dark soliton solutions, as given by
\begin{equation}
u (\eta, s) = u_0 \tanh (u_0 \eta) \exp (i u_0^2 s)
\label{5.11b} \end{equation}
where $u_0$ is a constant. Going back to the initial space time variables $(z, t)$, we can write the corresponding density perturbations as
\begin{equation}
\tilde n ({\bf r}, t) = A_0 F_{pl} ({\bf r}) \tanh [ u_0 ( k z - \omega t) ] \exp (i k z - \omega' t)
\label{5.12} \end{equation}
where we have introduced a new amplitude $A_0$, and defined the nonlinear mode frequency $\omega'$, as
\begin{equation}
A_0 = \sqrt{\frac{2}{W (\omega)}} \, \omega u_0 \; , \quad \omega' = \omega (1 - u_0^2)
\label{5.12b} \end{equation}
This shows that nonlinear twisted phonon modes, with dark soliton envelopes, can exist in a Bose Einstein condensate, which are different from the soliton structures usualy considered in a condensate \cite{phetick,book2}.
 
\section{Conclusions}

In this work, we have considered the existence of twisted phonon modes in a condensate. These are elementary excitations satisfying the usual Bogoliubov dispersion relation, but carrying a finite amount of angular momentum, or vorticity. They can be seen as phonon vortices, clearly distinct from the usual quantum vortices, which are static nonlinear structures with no dispersion relation. We have established the space and time solutions of these twisted modes, and determined the associated angular momentum. We have also shown that nonlinear twisted phonon structures can also exist, as described by the solutions of a defocusing NLS equation, similar to the GP equation of the condensate but now related to the phonon field.  New soliton solutions were shown to exist,  which correspond to envelope acoustic solitons, insensitive to the sign of the interaction parameter $g$.

We notice that the existence of such twisted modes in a condensate, described by both linear and nonlinear wave solutions, changes our traditional view of turbulence in a condensate, according to which vorticity was only carried by quantum vortices, and phonons could only carry linear momentum. A new paradigm is therefore needed, where quantum vortices at zero frequency coexist with phonon vortices with a broad frequency spectrum.

Finally, we would like to add that the paraxial wave approximation was only used here for practical reasons, and is not a limitation for the existence of phonons with finite vorticity. The existence of twisted phonons can indeed be equally demonstrated for non-paraxial wave geometries. In general, twisted modes could be described by expanding the three dimensional wave equation (\ref{3.1}) in spherical harmonics. This would lead to a similar but much heavier theoretical description of vortical phonons. The possible excitation of such vortical or twisted phonons by a potential perturbation moving through a condensate, similar to that recently considered by \cite{gammal}, and the relevance of nonlinear mode coupling phenomena, will described in a future publication.

\end{document}